\documentclass[express-paper, conference]{aesconf}

\graphicspath{{./}{figures/}}

\usepackage[utf8]{inputenc}

\usepackage{microtype}
\usepackage{tabularx}

\usepackage[numbers,square]{natbib}

\usepackage{booktabs}
\usepackage{color}
\usepackage{url}

\title{Beat-Based Rhythm Quantization of MIDI Performances}

\author[1]{Maximilian Wachter}
\author[1,2]{Sebastian Murgul}
\author[2]{Michael Heizmann}

\affil[1]{Klangio GmbH, Karlsruhe, Germany}
\affil[2]{Institute of Industrial Information Technology, Karlsruhe Institute of Technology, Karlsruhe, Germany}

\correspondence{Maximilian Wachter}{max.wachter@klang.io}

\lastnames{Wachter, Murgul, Heizmann}

\shorttitle{Audio algorithms and results}

\begin{document}

\twocolumn[
  \maketitle 

  \begin{onecolabstract}
    We propose a transformer-based rhythm quantization model that incorporates beat and downbeat information to quantize MIDI performances into metrically-aligned, human-readable scores. We propose a beat-based pre-processing method that transfers score and performance data into a unified token representation. We optimize our model architecture and data representation and train on piano and guitar performances. Our model exceeds state-of-the-art performance based on the MUSTER metric.
  \end{onecolabstract}
]

\section{Introduction}

This paper introduces a novel transformer-based approach for beat-based rhythm quantization of symbolic MIDI performances. Rhythm quantization aims to recover the intended notated rhythm from expressive performances, a fundamental task in music information retrieval and automatic music trancription. Our method uniquely integrates beat information and performance timing into a unified tokenized input representation, enabling the model to learn rhythmic structure by modeling timing deviations relative to an underlying metrical grid. Unlike most quantization models, our model is capable of leveraging metronome information, which entails the possibility of completely eliminating the uncertainty of beat estimations.

\section{Methods}

To support diverse musical contexts, we designed a flexible preprocessing pipeline that uses beat estimations or ground truth beats as a priori information and adapts to multiple time signatures without requiring explicit time signature tokens. This enables the model to generalize and successfully quantize rhythms in time signatures unseen during training. We propose a confusion-based evaluation metric tailored to beat-aligned quantization, incorporating both onset accuracy and note value correctness, which is better suited to evaluating timing on a score-level than frame-based metrics.

We used the T5 transformer for our model architecture, which we subsequently optimized in terms of efficiency and quality. For training our model we used the ASAP dataset \cite{foscarin_asap_2020}, a piano dataset containing performance MIDI files with accompanying scores as well as beat and downbeat annotations. We trained our model on $N$ measure sequences, assuming a one-to-one correspondence. Due to the sometimes poor alignment between scores and performances we filtered the dataset based on how well input and target sequences matched. We subsequently optimized the model based on input sequence length, note ordering, and data augmentation techniques such as pitch transposition, note deletion, and duration noise in order to maximize musical note onset accuracy.

\section{Experiments}
Our optimized checkpoint achieves an onset F1 score of 97.3\% on the ASAP dataset and a note value accuracy of  83.3\%.
Extending beyond piano, we adapted the model to guitar performances using the Leduc dataset, demonstrating that instrument-specific training yields improved quantization results. This highlights the importance of capturing instrument-dependent rhythmic interpretation characteristics for precise quantization.

Finally, we benchmarked our approach against state-of-the-art methods using the test and training splits defined in the ACPAS dataset for computing the MUSTER metric \cite{nakamura_towards_2018}. In Table \ref{tab:comparison} we compared the onset and offset error rates $\epsilon_\textit{onset}$ and $\epsilon_\textit{offset}$ to other state-of-the-art probabilistic and deep learning-based approaches. We showed that our model is able to surpass state-of-the-art models in $\epsilon_\textit{onset}$ by using beat annotations, while getting surpassed in $\epsilon_\textit{offset}$ only by \cite{beyer_end_2024}.

\begin{table}[h]
  \centering
  \caption{Comparison of onset-time ($\epsilon_\textit{onset}$) and offset-time ($\epsilon_\textit{offset}$) error rates using the MUSTER metric.}
  \begin{tabularx}{0.49\textwidth}{p{4cm}|>{\raggedleft\arraybackslash}X|>{\raggedleft\arraybackslash}X}
    \hline
    \textbf{Method}                                  & \multicolumn{1}{>{\centering\arraybackslash}X|}{\textbf{$\epsilon_\textit{onset}$}} & \multicolumn{1}{>{\centering\arraybackslash}X}{\textbf{$\epsilon_\textit{offset}$}} \\
    \hline 
    Neural Beat Tracking \cite{liu_performance_2022} & 68.28                                                                               & 54.11                                                                               \\
    End-to-End PM2S \cite{beyer_end_2024}            & 15.55                                                                               & \textbf{23.84}                                                                      \\
    HMMs (J-Pop) \cite{shibata_non-local_2021}       & 25.02                                                                               & 29.21                                                                               \\
    HMMs (classical) \cite{shibata_non-local_2021}   & 22.58                                                                               & 29.84                                                                               \\
    \hline
    \textbf{Our Model}                               & \textbf{12.30}                                                                      & 28.30                                                                               \\
    \hline
  \end{tabularx}
  \label{tab:comparison}
\end{table}

\section{Discussion}

Our results confirm that explicit beat information can yield significant improvements in onset quantization performance. However, offset accuracy remains challenging due to our model's current inability to detect 32nd notes. Compared to similar models, ours excels in scenarios where explicit beat information is available like in cases where performances are recorded to a metronome. If metronome information is not available the performance is however highly dependent on beat estimation quality.

\section{Summary}

Our findings suggest that transformer-based models with beat-aware tokenization offer a powerful framework for expressive rhythm quantization across instruments and meters. Future work will explore incorporating explicit time signature modeling, and expanding to irregular meters and note values, aiming for even greater generalization and musical fidelity.

\bibliographystyle{jaes}

\bibliography{lbdp_quantization_camera_ready}

\end{document}